\documentclass[twocolumn]{openjournal}

\usepackage[english]{babel}
\usepackage[utf8x]{inputenc}
\usepackage[T1]{fontenc}
\usepackage{aas_macros}

\usepackage{graphicx} 
\usepackage{amssymb}  %

\begin{document}

\title{The one that got away:\\A unique eclipse in the young brown dwarf Roque 12}

\email{as110@st-andrews.ac.uk}

\author{Aleks Scholz}
\affiliation{SUPA, School of Physics \& Astronomy, University of St Andrews, North Haugh, St Andrews, KY16 9SS, United Kingdom}

\author{Dirk Froebrich}
\affiliation{Centre for Astrophysics and Planetary Science, School of Physical Sciences, University of Kent, Canterbury CT2 7NH, UK}

\author{Koraljka Muzic}
\affiliation{CENTRA, Faculdade de Ci\^{e}ncias, Universidade de Lisboa, Ed. C8, Campo Grande, 1749-016 Lisboa, Portugal}

\author{Jochen Eisl{\"o}ffel}
\affiliation{Th{\"u}ringer Landessternwarte Tautenburg, Sternwarte 5, D-07778 Tautenburg, Germany}

\begin{abstract}
We report the discovery of a deep, singular eclipse of the bona fide brown dwarf Roque 12, a substellar member of the Pleiades. The eclipse was 0.65\,mag deep, lasted 1.3\,h, and was observed with two telescopes simultaneously in October 2002. No further eclipse was recorded, despite continuous monitoring with Kepler/K2 over 70\,d in 2015. There is tentative ($2\sigma$) evidence for radial velocity variations of 5\,kms$^{-1}$, over timescales of three months. The best explanation for the eclipse is the presence of a companion on an eccentric orbit. The observations constrain the eccentricity to $e>0.5$, the period to $P>70\,d$, and the mass of the companion to $\sim 0.001-0.04\,M_{\odot}$. In principle it is also possible that the eclipse is caused by circum-sub-stellar material. Future data releases by Gaia and later LSST as well as improved radial velocity constraints may be able to unambiguously confirm the presence of the companion. This would turn the system into one of the very few known eclipsing binary brown dwarfs with known age.
\end{abstract}

\section{Introduction}

Stellar variability encodes information about stars and their environment. Firstly, specific types of variability can be used to infer fundamental properties of stars that are notoriously difficult to determine by other means (luminosity, radius, distance, rotation period). Second, variability can reveal the existence of unseen objects (e.g. exoplanets, stellar companions) or may be used to map unresolved environments (e.g., protoplanetary disks, AGN disks). And third, variability allows us to directly witness dynamic, stochastic processes in the surroundings of stars and other unresolved sources, including magnetic activity, weather, accretion flicker, or disk instabilities \citep{2011uvs..book.....P}.

Eclipses are a rich source of information as well as essential calibrators for stellar models. Among eclipsing binaries, those who belong to a cluster and thus have a known age deserve a special distinction. For brown dwarfs with masses below the sub-stellar limit, the lack of known eclipsing binaries at different ages has been a major hindrance for the development of models. Unlike stars, brown dwarfs never reach a steady state like the main sequence and cool down as they get older. The atmospheres of brown dwarfs transition from an initial state that is comparable to very low-mass stars to exoplanet analogues with complex molecular chemistry, dust formation, and clouds \citep{2014A&ARv..22...80H}. The deep convection zones as well as the decline of fusion processes in the core also pose challenges for models. Calibrators are therefore needed at all evolutionary stages. 

For a decade, only one bona fide brown dwarf eclipsing binary has been known, a member of the 1\,Myr old Orion Nebula Cluster \citep{2006Natur.440..311S}. This one has shown a temperature reversal, with the more massive component being cooler than the secondary and demonstrated the need to take into account the effects of magnetic fields in evolutionary models for brown dwarfs \citep{2007ApJ...664.1154S}. Two more eclipsing systems with a pair of brown dwarfs have been found more recently, both in systems with more components, one in the 5-10\,Myr Upper Scorpius association \citep{2016ApJ...816...21D} (but see also \citet{2018ApJ...865..141W}), and one with a grazing eclipse in the 45\,Myr old Argos moving group \citep{2020NatAs.tmp...43T}. A few more detached very low mass eclipsing binaries with a component below the brown dwarf mass limit have been discovered as well \citep[e.g.][]{2015A&A...584A.128L,2010ApJ...718.1353I,2011ApJ...730...79J}. For a large part of the age-mass parameter space, however, there is still no eclipsing sub-stellar system to provide a sanity check for evolutionary models.

Not only compact objects can cause eclipses. Stars (or brown dwarfs) are occasionally obscured by material in orbit around them. These sources are excellent laboratories to obtain spatial information about processes in disks or clouds or tori that are otherwise impossible to resolve. Stars like UX\,Ori \citep{1994A&A...292..165G}, KH15D \citep{2002PASP..114.1167H}, $\epsilon$\,Aur \citep{1971Ap&SS..10..332K}, AA\,Tau \citep{1999A&A...349..619B}, and RW\,Aur \citep{2016MNRAS.463.4459B} are prominent examples of systems with deep eclipses that have provided valuable insights into early stellar evolution and planet forming processes. For young low-mass stars, in particular, eclipses by circumstellar material may currently be the only passable method to gain spatial information on scales smaller than 1\,AU which cannot be reached with submm/mm interferometry or Adaptive Optics \citep{2020MNRAS.493..184E}. Moreover, the resolution that can be achieved by modeling variability caused by obscurations is independent of distance and applicable down to very faint stars. For brown dwarfs this technique is currently the only hope of mapping the planet-forming zone around them.

For all these reasons it is thought of as critical to find rare systems undergoing eclipses, even more so when the primary source of light is a brown dwarf. Here we report such a system, for which we observed one singular deep and short eclipse in October 2002, an event that is best explained by a sub-stellar/planetary companion on an eccentric orbit, but could in principle also be caused by circum-sub-stellar dust. 

\section{Roque 12}

\begin{table}[t]
\caption{Observational properties of Roque 12.}
\label{table1}
\begin{tabular}{lll}
\noalign{\smallskip}
\hline
\noalign{\smallskip}
property   & value & source \\
\noalign{\smallskip}
\hline
\noalign{\smallskip}
photometry: &      & \\
G-mag & $20.21\pm 0.01$ & Gaia DR2 \\
I-mag & $18.3 \pm 0.04$ & \citet{2000MNRAS.313..347P}\\
      & $18.5$          & \citet{1998ApJ...507L..41M}\\
Z-mag & $17.66\pm0.02$         & UKIDSS DR9 \\
Y-mag & $16.67\pm0.01$         & UKIDSS DR9 \\
J-mag & $16.12 \pm 0.09$ & 2MASS \\
      & $15.93 \pm 0.01$ & UKIDSS DR9 \\
H-mag & $15.38 \pm 0.12$ & 2MASS \\
      & $15.37 \pm 0.01$ & UKIDSS DR9 \\
K-mag & $14.72 \pm 0.09$ & 2MASS \\
      & $14.94 \pm 0.01$ & UKIDSS DR9 \\
W1mag & $14.72 \pm 0.03$ & AllWISE \\
W2mag & $14.62 \pm 0.06$ & AllWISE \\
\noalign{\smallskip}
\hline
\noalign{\smallskip}
kinematics: & & \\
parallax    & $8.6\pm 1.2$\,mas          & Gaia DR2\\
PM $\alpha$ & $18\pm 3\,$mas\,yr$^{-1}$  & Gaia DR2\\
PM $\delta$ & $-44\pm 2$\,mas\,yr$^{-1}$ & Gaia DR2\\
RV          & $2.6\pm 2.3$\,kms$^{-1}$   & this paper \\
\noalign{\smallskip}
\hline
\noalign{\smallskip}
SpT & M7.5  & \citet{1998ApJ...507L..41M}\\
\noalign{\smallskip}
\hline
\noalign{\smallskip}
\vspace{0.1cm}
\end{tabular}
\end{table}

\begin{figure}
\center
\includegraphics[width=1.0\columnwidth]{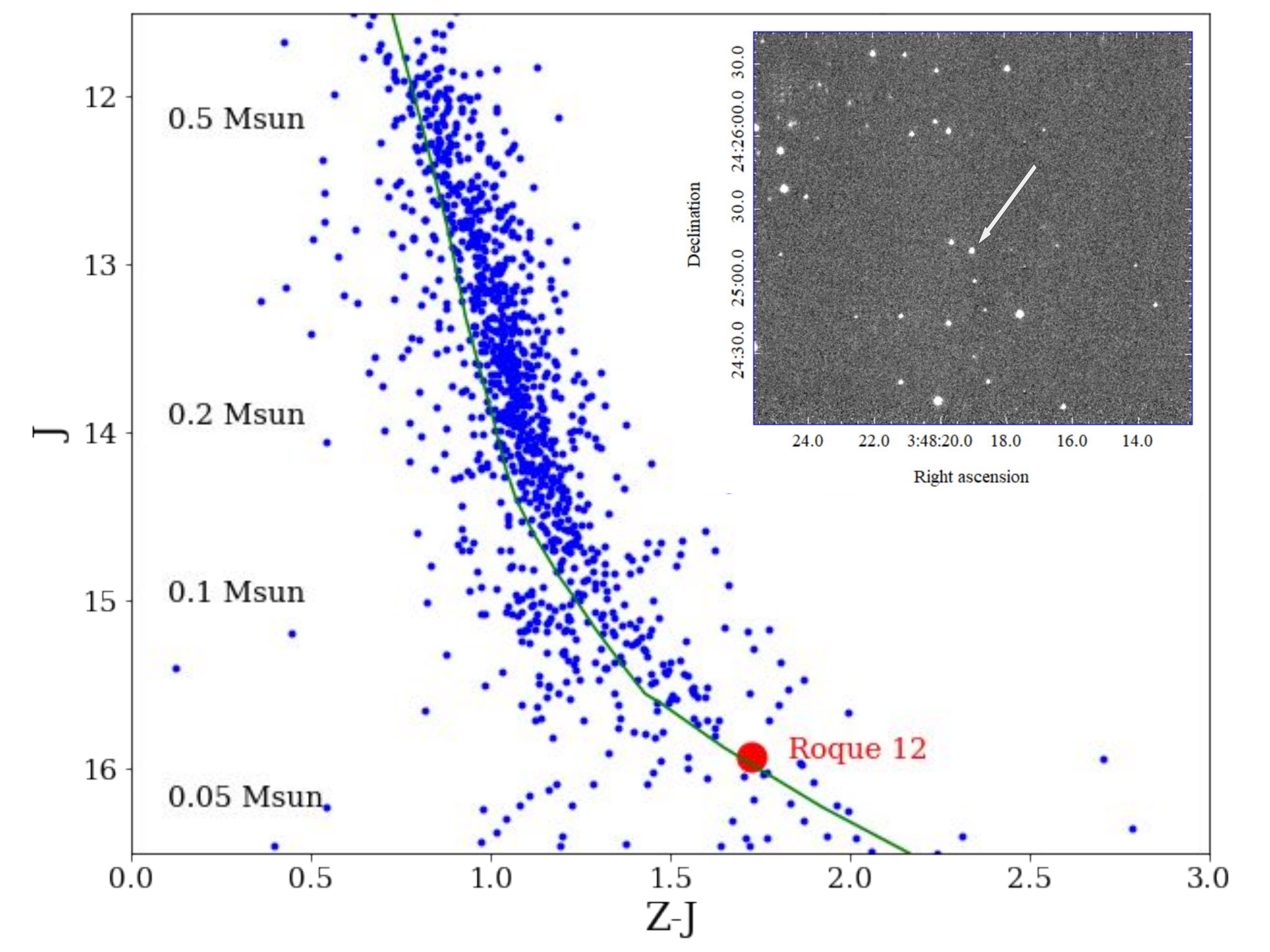}
\caption{Colour-magnitude diagram from the UKIDSS/GCS photometry published by \citet{2012MNRAS.422.1495L}. The position of R12 is marked. A 120\,Myr model isochrone (assuming a distance of 134\,pc) with mass scale is overplotted \citep{2015A&A...577A..42B}. The inset shows a $3\times3$\,arcmin finding chart for R12, from the UKIDSS J-band image.
\label{fc_cmd}}
\vspace{0.3cm}
\end{figure}

Roque 12 is a bona fide brown dwarf in the Pleiades star cluster, in the following called R12 (also named BPL172; $\alpha$ 03:48:19.0 $\delta$ +24:25:13.0 J2000.0). It was identified in several deep wide-field surveys of the Pleiades from colour-magnitude diagrams \citep{1997ApJ...491L..81Z,2000MNRAS.313..347P}. To our knowledge, the only spectrum in the literature was published by \citet{1998ApJ...507L..41M}. The object has a spectral type of M7.5 and shows H$\alpha$ emission (equivalent width 19.7\,\AA). No Li\,I absorption at 6707\,\AA\ is reported in \citet{1998ApJ...507L..41M}, with an upper limit in the EW of 1.5\,\AA. Compared to the Li\,I EW in other Pleiades brown dwarfs (0.5-2.5\,\AA), this does not firmly rule out the presence of Li\,I. The kinematics, colours, and H$\alpha$ emission are all consistent with cluster membership \citep{1998ApJ...507L..41M,2000MNRAS.313..347P,2007ApJS..172..663S,2012MNRAS.422.1495L}. 

In Fig. \ref{fc_cmd} we show the position of R12 in the colour-magnitude diagram of Pleiades members, as published by \citet{2012MNRAS.422.1495L}. Judged by its position relative to the sequence of cluster members and the theoretical isochrone, R12 is a single object with a mass of around 0.06\,$M_{\odot}$. The figure also includes a finding chart from the J-band image of UKIDSS/GCS On the sky, the object sits in an uncrowded area. The brown dwarf has an entry in the Gaia Data Release 2 catalogue \citep{2018A&A...616A...1G}, with a parallax of $8.6\pm 1.2$\,mas and proper motions of $\mu_\alpha = 18\pm 3\,$mas\,yr$^{-1}$, $\mu_\delta = -44\pm 2$\,mas\,yr$^{-1}$. These values are again in line with what we expect for members of the Pleiades. Based on Gaia DR2, \citet{2019PASP..131d4101G} derive a membership probability of 0.956. The object is also listed in AllWISE \citep{2013yCat.2328....0C}, with magnitudes of 14.7 and 14.6 at 3.4 and 4.6\,$\mu m$, respectively, that means, there is no evidence for warm circum-sub-stellar dust. Photometric and kinematic properties are summarised in Table \ref{table1}.

\section{Observations}

\subsection{Calar Alto 1.23m}

In October 2002 we monitored a field in the Pleiades covering R12 over three weeks, using the 1.23\,m telescope at the German Spanish Astronomical Centre on Calar Alto. The telescope was equipped with a 2k$\times$2k CCD covering $17'\times17'$ on the sky. All images were taken in the I-band with 600\,sec exposure time. Observations were carried out on 15 nights between Oct 2 and Oct 18; several nights of the run were not usable due to unfavourable weather conditions. The main purpose of this observing run was to measure rotation periods for very low mass Pleiades members; the full details of analysis and results are reported in \citet{2004A&A...421..259S}. Light curves were derived using the 'Optimal Image Subtraction' method \citep{1998ApJ...503..325A} in an implementation developed by the Wendelstein Calar Alto Pixellensing Project \citep{2001A&A...379..362R,2002A&A...381.1095G}. We achieved a typical photometric precision of 1\% for stars with $I=17$\,mag and 2\% for stars at $I=18$. For R12, the standard deviation of the lightcurve outside the eclipse is 0.034\,mag.

\begin{figure*}
\center
\includegraphics[width=1.0\columnwidth]{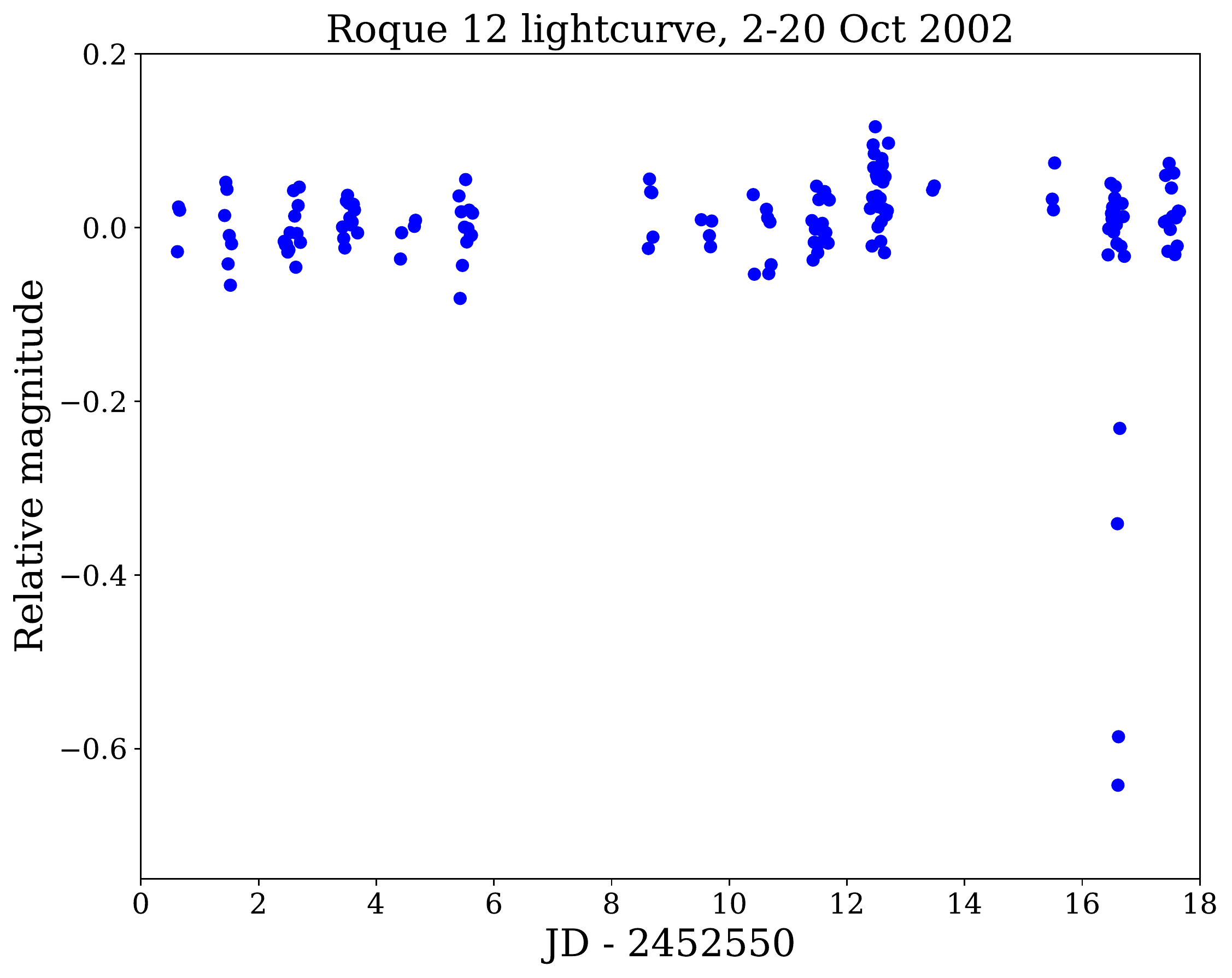}
\includegraphics[width=1.0\columnwidth]{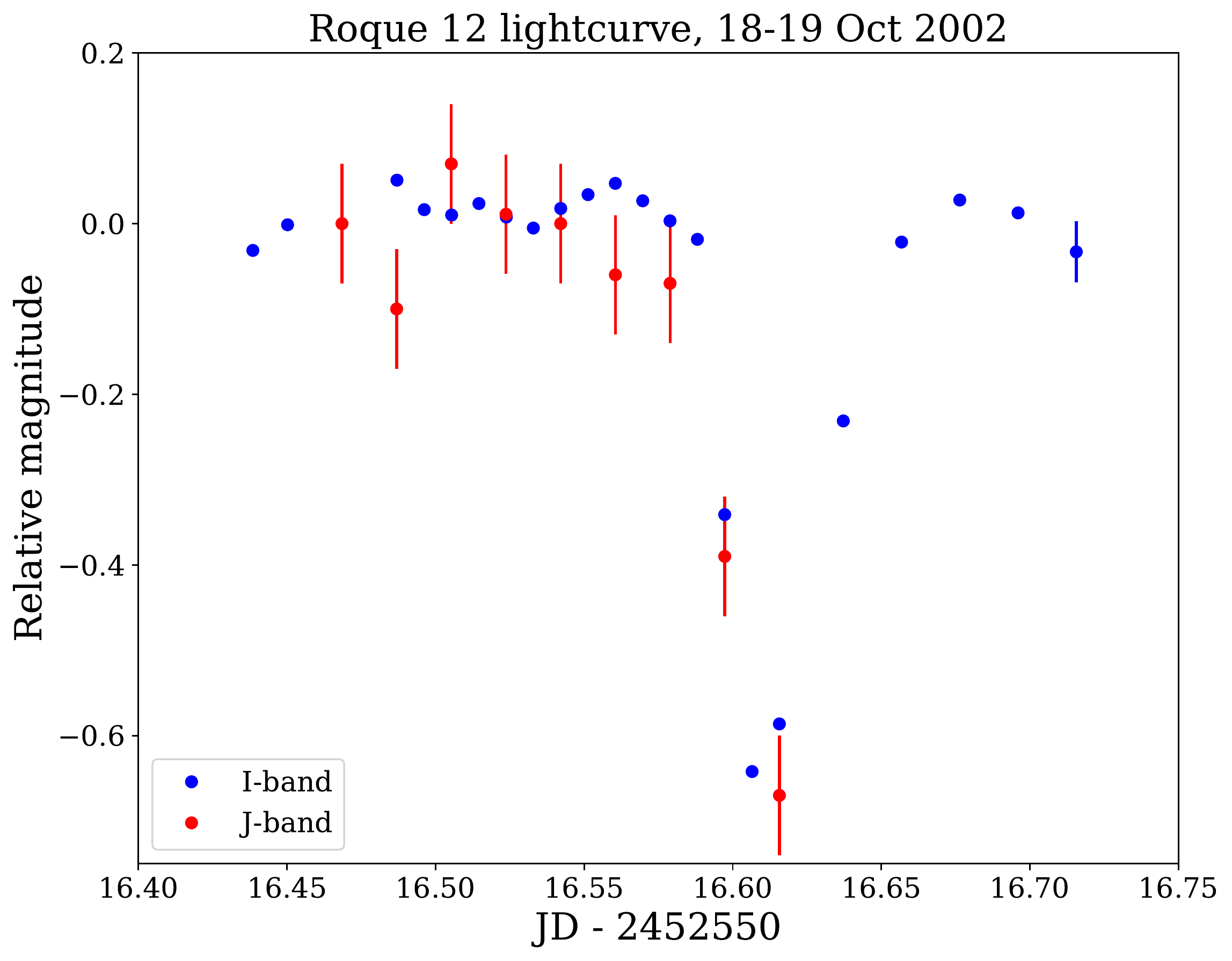}
\caption{{\bf Left:} I-band lightcurve for R12 for Oct 2-20 2002, observed with the 1.23\,m telescope on Calar Alto. The eclipse is clearly detected at $t=16.6$\,d. {\bf Right:} I- and J-band lightcurve for the night Oct 18-19 2002, observed with the 1.23\,m and 2.2\,m telescopes on Calar Alto. Typical errorbars are indicated. Within the errors, the eclipse depth is the same in the two bands. 
\label{lc}}
\vspace{0.3cm}
\end{figure*}

The full lightcurve for R12 is shown in the left panel of Fig. \ref{lc}, with a closeup view provided in the right panel. In the penultimate night of the observing run a deep eclipse in R12 is visible. The event is clearly detected in photometry from the reduced frames at $t=16.61$\,d ($JD = 2452566.61$\,d). It is also visible in the lightcurve derived from raw images. The eclipse reaches a depth of 0.65\,mag compared with the average I-band magnitude over the entire campaign. The length of the event, which is covered by only four data points, is $1.3\pm0.3$\,h.

\subsection{Calar Alto 2.2m}

During the last part of this campaign (Oct 17-19), we observed the fields discussed above also with the 2.2\,m telescope on Calar Alto using the infrared camera MAGIC \citep{1993SPIE.1946..605H}. This camera in wide-field mode gives a field of view of $7'\times 7'$. We observed alternately in J- and H-band, with exposures times of $13\times5$ and $19\times5$\,sec. This observing sequence covered half of the eclipse seen in the I-band data in the neighbouring 1.23\,m telescope, but only in the J-band. A standard data reduction was carried out, using sky subtraction, bias correction, and flatfield correction. The lightcurve for R12 was derived using standard aperture photometry followed by differential correction using non-variable stars in the same field. The full details of this campaign are described in \citet{2005A&A...438..675S}. Based on Fig. 1 of this paper, the typical uncertainty for the J-band data of R12 is $\sim 0.07$\,mag. The eclipse lightcurve from this run together with the close-up view of the I-band lightcurve is shown in Fig. \ref{lc}, right panel. The eclipse depth in the J-band is the same as in the I-band, within the uncertainties. 

\subsection{Kepler K2}
\label{k2}

\begin{figure}
\center
\includegraphics[width=1.0\columnwidth]{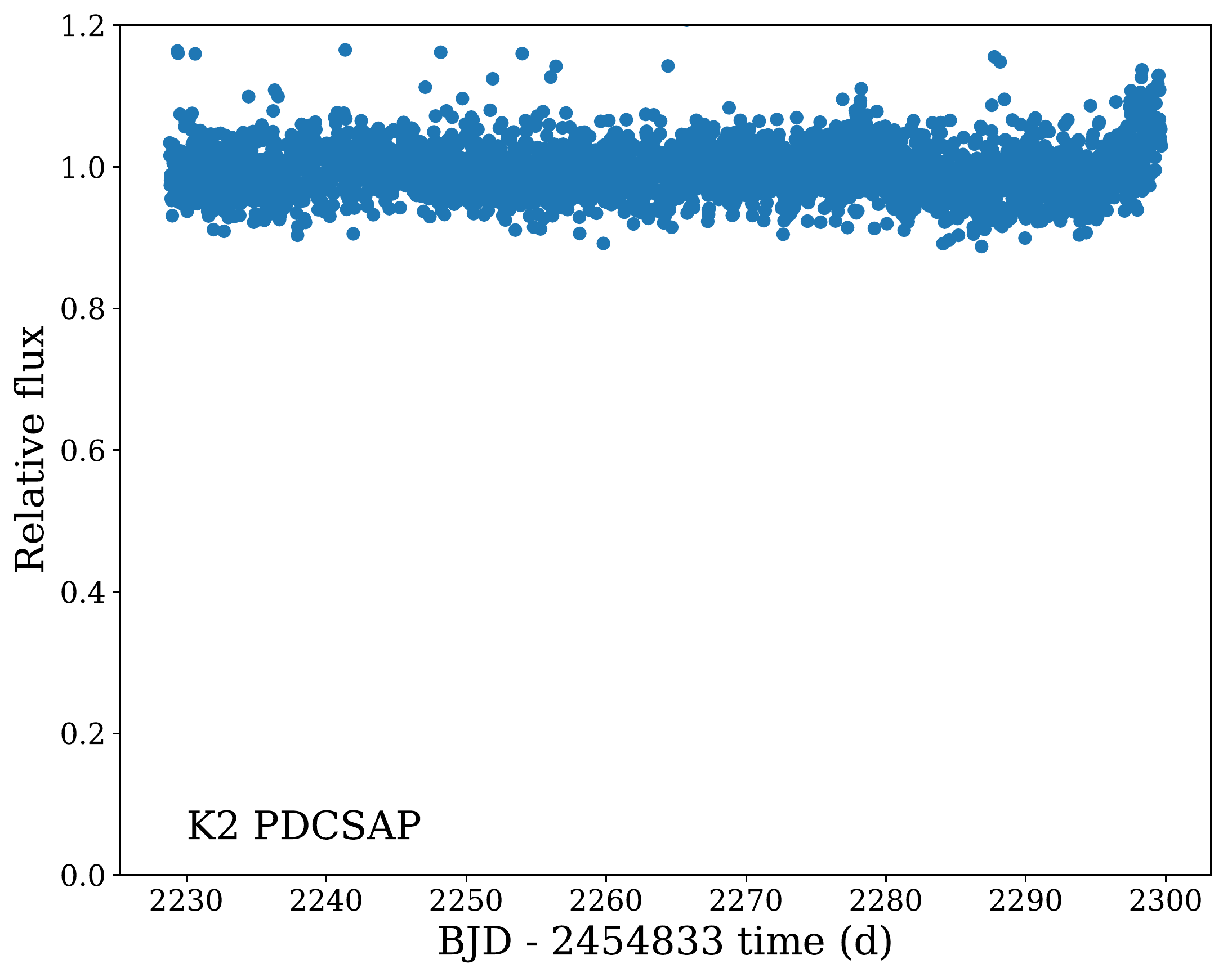}
\caption{K2 lightcurve for Roque 12. Shown are the relative fluxes from the PDCSAP dataset. \vspace{0.3cm}
\label{k2lc}}
\end{figure}

Roque 12 was observed as part of campaign 4 of the Kepler K2 mission \citep{2014PASP..126..398H}, under EPIC no. 211090981, in GO program 4026 (PI: A. Scholz). Continuous observations with a standard cadence of 30\,min were carried out from Feb 07 2015 to Apr 23 2015. The Kepler magnitude for R12 is 17.628, at the faint end of the magnitude distribution of K2 targets. We retrieved the lightcurve for R12 from MAST on September 4th 2015. The PDCSAP lightcurve after removing systematics \citep{2010SPIE.7740E..1UT,2010ApJ...713L..87J} is shown in Fig. \ref{k2lc}, after dividing by the mean.

The lightcurve contains 3470 data points in total, 190 of them have relative fluxes $>1.2$ and are not shown in the figure. No data points are below 0.8, however. The standard deviation of the lightcurve without the positive outliers is 0.035, the average photometric error 0.024. Any repetition of the eclipse observed in 2002 would be clearly visible in the lightcurve as multiple negative outliers. Thus, if the eclipses are periodically recurring, the K2 lightcurve puts a definite lower limit of 70\,d on the period. More recent versions of the same lightcurve, after correcting for systematics, do not give any different result. To rule out that the eclipse make the object too faint to be detected, we investigated the K2 lightcurve interactively with the {\tt Lightkurve} tool \citep{2018ascl.soft12013L}. The source, however, is visible in all frames, including the ones with 'nan' data points in the lightcurve, without any evidence for anomalous behaviour.

\subsection{Spectroscopy with VLT-UVES}
\label{rv}

In 2003-2004, a total of 13 epochs of high-resolution spectroscopy were obtained with UVES at the ESO-VLT, in the framework of ESO program 072.C-0071. In principle, this data set is useful in constraining the presence of companions through measurement of radial velocities (RV). The observations were carried out 6-11th October 2003 (4 epochs) and 13-28th January 2004 (9 epochs). These spectra span a wavelength range from 665 to 1042\,nm with a nominal resolution of 34000. Only the red arm was used, with grism 'CD\#4'. The integration time for each spectrum was 3000\,sec. 

We retrieved the extracted spectra for this run from the ESO archive in April 2020; the pipeline reduction was performed on 2017-10-26. The signal-to-noise ratio in the spectra is poor; it ranges from $\sim 1$ at the blue end to $\sim 10$ at the red end. As a result, only the strongest features are clearly detected and the constraints on RV are weak. We note, in particular, that the LiI line at 6707\,\AA\ is not detected.

\begin{figure*}
\center
\includegraphics[width=1.0\columnwidth]{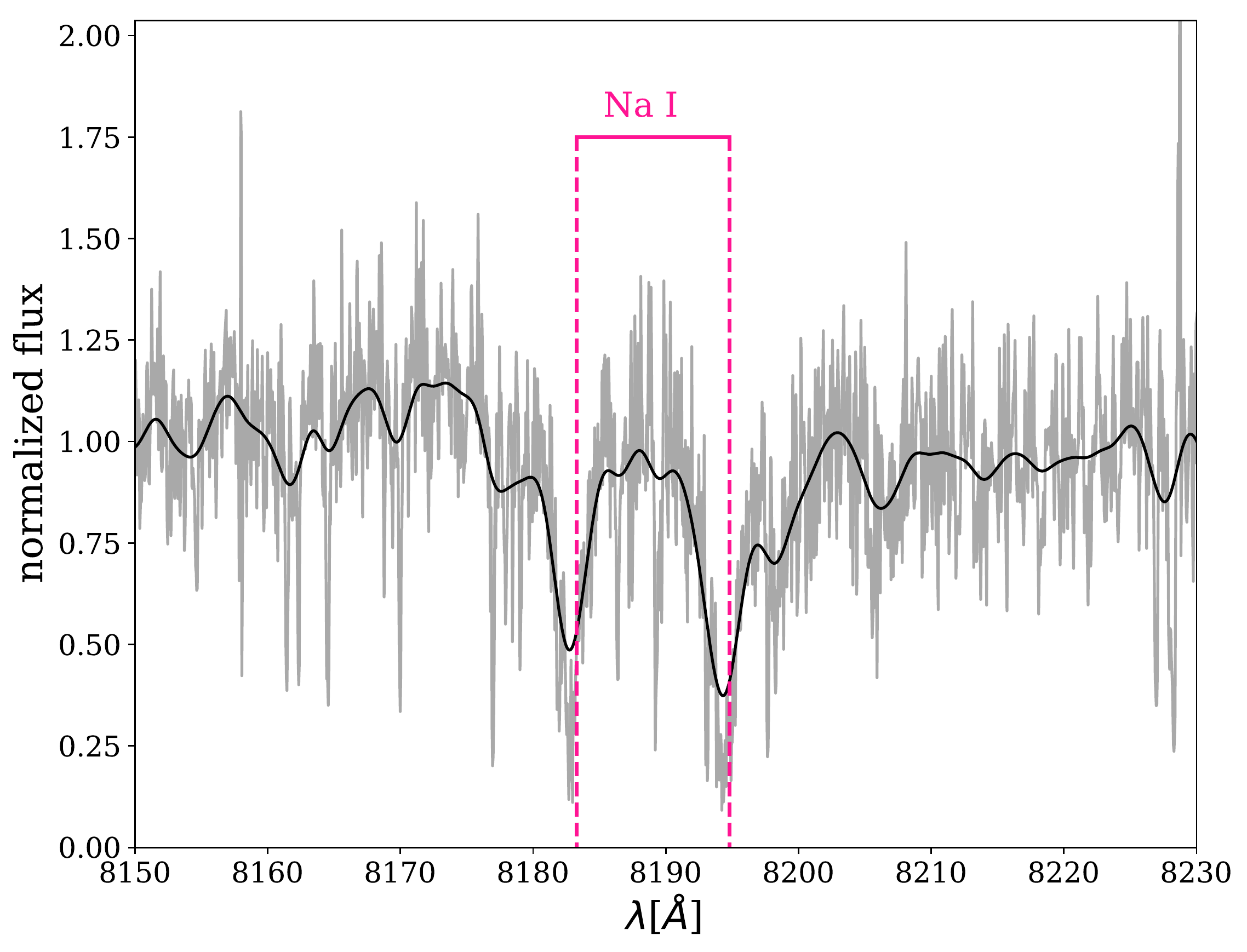}
\includegraphics[width=1.0\columnwidth]{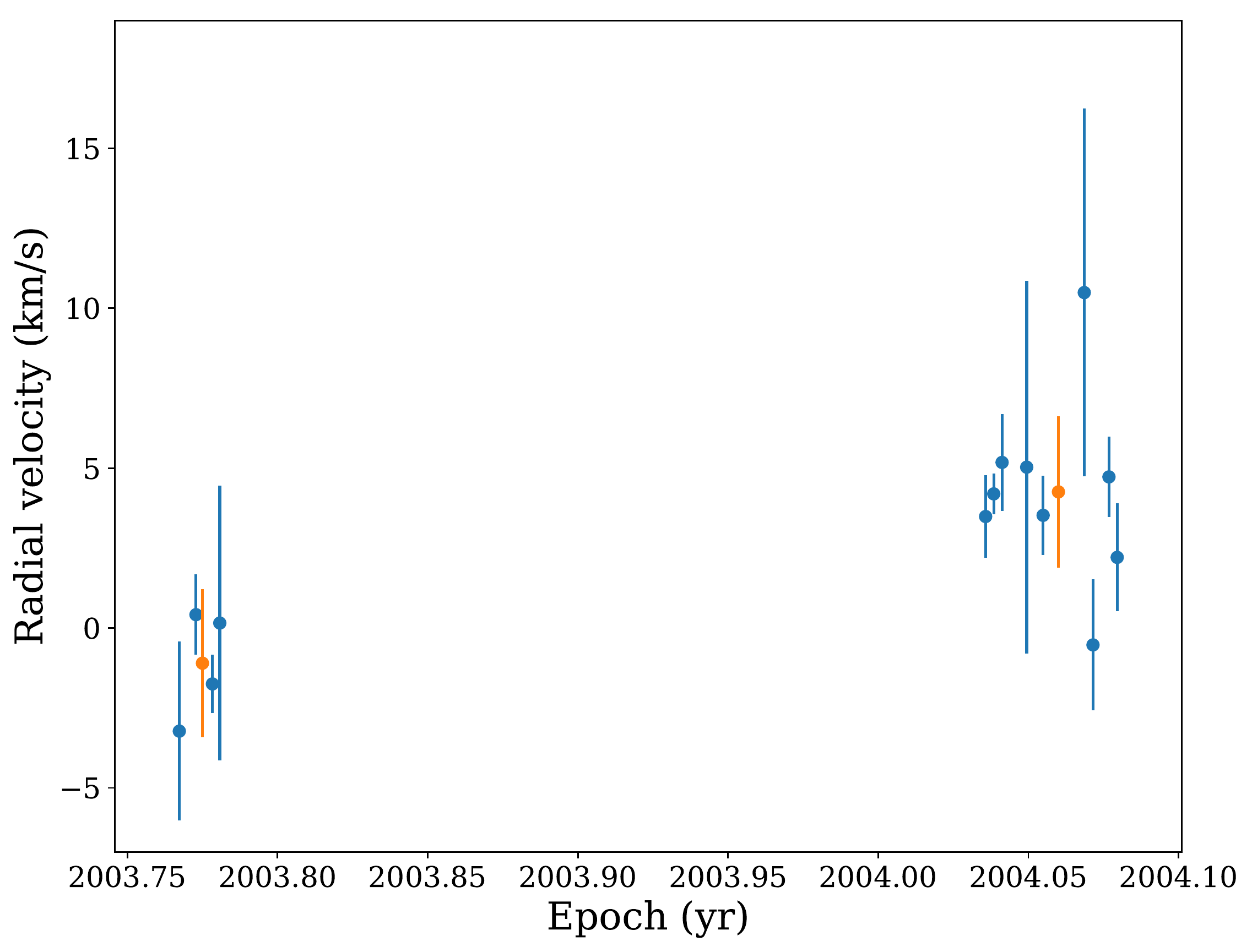}
\caption{{\bf Left:} Example spectrum obtained with UVES at ESO/VLT on October 8 2003. The grey line is the reduced spectrum, the black line the same convolved with a Gaussian with a standard deviation of 0.8\,\AA. The rest-frame wavelengths of the Na doublet used for measuring radial velocities are marked. {\bf Right:} Radial velocity measurements over time (blue) and the averages for the two groups of epochs (orange).} \vspace{0.3cm}
\label{rvfig}
\end{figure*}

We measured RV for all epochs by fitting the Na doublet at 8183/8195\,\AA, the strongest absorption lines in the spectrum of R12, with a Gaussian. A small part of an example spectrum with these two lines is shown in the left panel of Fig. \ref{rvfig}. We adopted the average of the two measurements as radial velocity, and half of the difference between the two as uncertainty. The resulting RV after heliocentric correction are shown in the right panel of Fig. \ref{rvfig}.

Radial velocities range from $-3$ to $10$\,kms$^{-1}$, with an average of $2.6\pm 2.3$\,kms$^{-1}$. For comparison, the RV of the Pleiades cluster is $\sim 6$\,kms$^{-1}$ \citep{2009A&A...498..949M}. The average RV of R12 is $-1.1\pm 2.3$\,kms$^{-1}$ in October 2003 and $4.3\pm 2.4$\,kms$^{-1}$ in January 2004. This might indicate long-term variations in the radial velocity with an amplitude of $\sim 5$\,kms$^{-1}$, but at this point the evidence is only tentative. 

\subsection{Summary of observations}

The unique event seen in October 2002 with two Calar Alto telescopes simultaneously remains to date the only eclipse observed in this object. The non-detection in the Kepler/K2 lightcurve limits any possible period to $>70$\,d. There is weak evidence for radial velocity variations on timescales of months. We note that to our knowledge there is no additional indication of photometric variability in the literature or in public archives. 

\section{Explaining the eclipse}
\label{discussion}

The possible explanations for the singular eclipse fall into three broad categories: a sub-stellar or planetary companion to R12, a cloud in orbit around R12, an object between Earth and the Pleiades, but not associated with R12. These options will be discussed in the following. We do not claim that the discussed ideas exhaust the entire range of plausible explanations. 

\subsection{Companion}
\label{companion}

If the eclipse is caused by a companion, the characteristics of the eclipse give us constraints on its nature. The lack of a measurable colour change as well as the fact that R12 is positioned close to the single object sequence in Pleiades colour-magnitude diagrams \citep{2012MNRAS.422.1495L}, tells us that the second body contributes little to the flux ($\lesssim 10$\%). Based on the 120\,Myr DUSTY isochrone \citep{2001ApJ...556..357A}, this limits the secondary mass to $\lesssim 0.04\,M_{\odot}$, i.e. it cannot be an equal mass companion. The eclipse depth of 0.65\,mag means that the eclipsing body, if it contributes no flux, has to cover almost half (45\%) of the primary object. Taken together, the companion can only be a low-mass brown dwarf or a giant planet; both types of object would have sizes comparable to the primary brown dwarf.

\begin{figure*}
\center
\includegraphics[width=1.0\columnwidth]{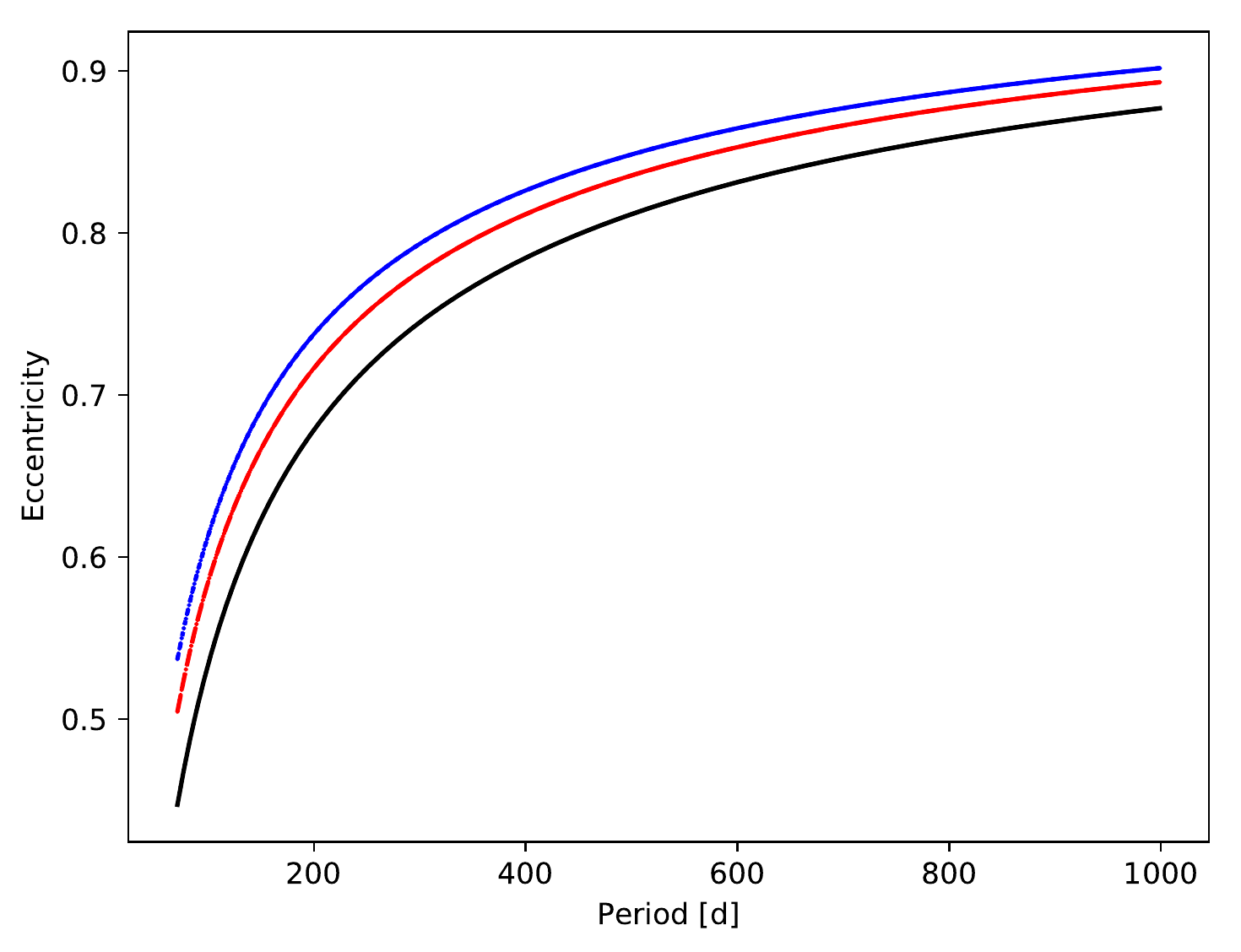}\hfill
\includegraphics[width=1.0\columnwidth]{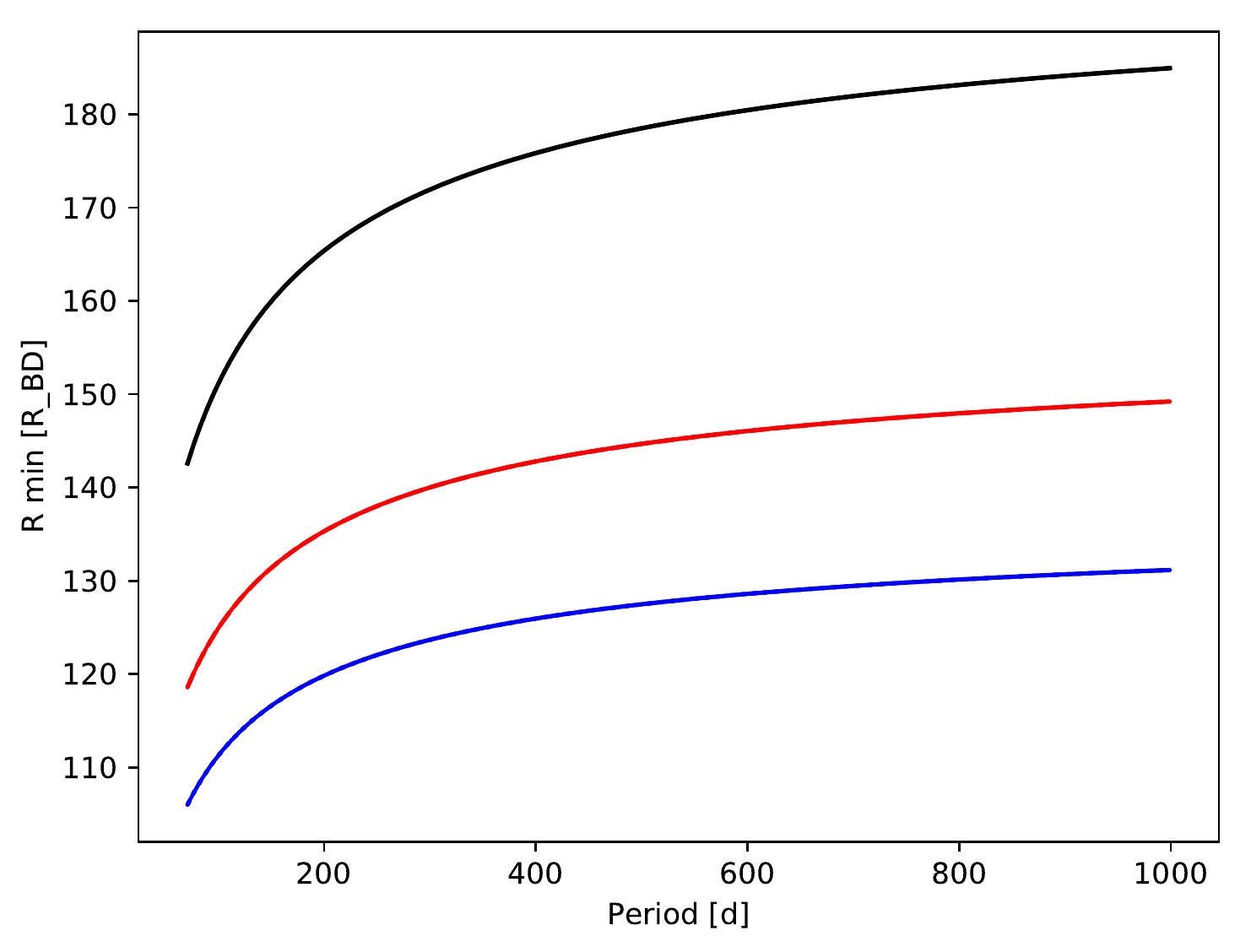}
\caption{{\bf Left:} Eccentricity as function of period for a 1.3\,hr long eclipse of a 0.06$\,M_{\odot}$ brown dwarf by a 0.04 (black), 0.02 (red), 0.01$\,M_{\odot}$ (blue) companion, assuming an object radius of 0.13$\,R_{\odot}$ and the eclipse occurring at periastron. {\bf Right:} Minimum distance between the two objects in units of the object radius (0.13$\,R_{\odot}$). \label{bd_ecc}}
\vspace{0.3cm}
\end{figure*}

For the observed eclipse duration of 1.3\,h, the expected period for a circular Keplerian orbit is $\sim 15$\,days, with large uncertainty given the unknown mass of the companion and the poorly sampled eclipse. Periods shorter than that can be ruled out, based solely on the eclipse duration. As stated in Sect. \ref{k2}, periods $<70$\,d are unambiguously excluded by the K2 lightcurve. 

For an eccentric orbit to produce such a short eclipse in a system as described, the eccentricity has to be at least $e\sim 0.5$. The required eccentricity is shown in Fig. \ref{bd_ecc}, left panel, as a function of period for three assumed companion masses, derived from Kepler's laws. With object masses $M_1$, $M_2$, semi major axis $a$, the eclipse duration $t$, the object radius $R$, we plot the eccentricity $e$ as:

\begin{equation}
e = \frac{X-1}{X+1} 
\end{equation}
with
\begin{equation}
X = \left( \frac{2R}{t} \right) ^2 \frac{a}{G(M_1+M_2)}
\end{equation}

For this calculation we assumed that the eclipse happened at periastron, the time of the closest approach; if it happened at any other point in the orbit, the eccentricity would need to be larger. With a highly eccentric orbit, the period may be in the range of several years. The resulting minimum orbital separations are plotted in the right panel of the same figure, again as a function of period. Values of 120-180 times the radius of the brown dwarf are plausible, corresponding to 0.07-0.11\,AU.


In an eccentric Keplerian orbit, companions with masses of $M_B = 0.01-0.04\,M_{\odot}$ would cause an orbital velocity in the primary of 6-15\,kms$^{-1}$ at periastron, and a few kms$^{-1}$ at apastron; the latter will depend on the orbital period. These values are consistent with the tentative radial velocity variations reported in Sect. \ref{rv}.

Physically, the described system is plausible, but how unusual would it be? The fraction of unresolved brown dwarf-brown dwarf systems in the Pleiades is $27.1\pm 5.8$\% \citep{2012MNRAS.422.1495L}. Giant planets around brown dwarfs have also been found, from direct imaging \citep{2004A&A...425L..29C} and microlensing \citep{2013ApJ...778...38H,2018AJ....155..219J}. Thus, finding an eclipsing brown dwarf or giant planet companion around a brown dwarf is not implausible.

High eccentricities as needed to explain the singular eclipse for R12 have also been observed for similar and related systems. \citet{2011ApJ...733..122D} investigated binaries among very low mass stars and brown dwarfs and find that the eccentricities span a broad range, including those with high eccentricities up to $e \sim 0.8$. Their sample covers a very wide range in periods, from a few days to many years, including the plausible periods for the R12 system. 

In a population level analysis of directly imaged sub-stellar and planetary companions around stars, \citet{2020AJ....159...63B} find that the eccentricity distribution is flat between 5 and 100\,AU separations. Brown dwarf companions exhibit a broad peak between $e=0.6-0.9$. High eccentricities $>0.5$ are not very common among the known giant planets in close orbit around stars, but they do exist in significant numbers \citep{2015ARA&A..53..409W}. 
 
If R12 had a low-mass companion, it would likely have avoided circularisation. For low-mass stars, the cut-off period for tidal circularisation of orbits at the age of the Pleiades is around 8\,d \citep{2005ASPC..333....4Z}. This value should depend weakly on the stellar mass/radius and is expected to be slightly larger for very low mass systems. The period in the putative R12 system, however, is sufficiently long to be above this threshold. 
 
We conclude that an ultra-low mass companion on an eccentric orbit is a plausible explanation for the eclipse observed for R12. 

\subsection{Circum-sub-stellar clouds}

In this scenario, we observed a dusty, circum-sub-stellar, optically thin cloud passing in front of the brown dwarf. Due to the lack of reddening observed during the eclipse, any optically thin occulter has to be made of grains that are large in comparison with the wavelength, i.e. $>>1\,\mu m$, to cause grey extinction. Grains of that nature are readily formed in accretion disks and should also be common in debris disks \citep{2014prpl.conf..339T}. 

The fast ingress and egress would point to a cloud moving at a high velocity of $\sim 50$\,kms$^{-1}$, to be able to travel the diameter of the brown dwarf within an hour. The uniqueness of the eclipse could be explained by material spiralling or falling into the central object. As shown in \citet{2019MNRAS.484.4260S}, only very little dust ($<<M_{\mathrm{Earth}}$) is needed to cause substantial absorption along the line of sight. That means, this explanation does not necessarily conflict with the lack of infrared excess. A possible scenario to create such a cloud would be a something akin to a comet swarm on a highly eccentric orbit being destroyed as it approaches the central object. 

There is a growing class of evolved stars showing eclipses or 'dips' in the lightcurve due to circumstellar material, although there is no IR excess indicating the presence of a debris or accretion disk. The most prominent one might be Boyajian's Star found in the Kepler data set \citep{2016MNRAS.457.3988B}, but there are other examples at younger ages \citep{2017AJ....153..152S}, see also \citet{2019MNRAS.484.4260S} for a discussion. A brown dwarf in $\sigma$\,Orionis might also belong into that category \citep{2017A&A...608A..66E}. Among Pleiades members, a few examples of such disk-less 'dippers' have been identified \citep{2016AJ....152..114R}. R12 could be another enigmatic member of this family of objects, with one singular deep eclipse, instead of many shallow ones. We note that the radial velocity variations, if confirmed, cannot be explained by dust occultations. 

\subsection{Occulter unrelated to R12}

This scenario can be sub-divided in two possibilities, depending on the location of the occulter. One option is that R12 is eclipsed by an object in the solar system. R12 is a point source and at the minimum of the eclipse the flux is still more than 50\% of the light out of eclipse. Therefore, the occulter should have been visible in our images before and after the eclipse, which is not the case. Asteroid occultations are also very short (seconds for main belt asteroids or minutes for Kuiper belt objects), not comparable to the observed eclipse in R12. Therefore, this option can be excluded.

The alternative possibility is the presence of an occulter along the line of sight, for example, a dusty cloud. For typical ISM grains, the eclipse in the I-band should be about 1.7 times deeper than in the J-band \citep{1990ARA&A..28...37M}, which is not observed. In theory, the occulter could be part of a disk with processed grains around a fast moving young star in the foreground. There is no evidence of such an object in our images or in the archives. The proximity of R12 as member of the Pleiades renders this explanation implausible. We refer to \citet{2016ApJ...829L...3W} for a more detailed discussion of this kind of scenario. 

Instead of an interstellar dust cloud, the occulter could be a free-floating low-mass brown dwarf or giant planet passing through the line of sight. Such objects can be sufficiently red and faint to be undetectable in our optical images. From surveys of star forming regions \citep{2012ApJ...756...24S} and from microlensing studies \citep{2017Natur.548..183M} we know that these type of objects are significantly less common than stars in the Milky Way, but still exist in substantial numbers. For an eclipse of this nature to happen, the free-floating planet or brown dwarf would have to be located in the cone-shaped volume towards R12. Typical space densities for this type of object are in the range of 0.1\,pc$^{-3}$ \citep{2015MNRAS.449.3651M}. With this number, the volume towards R12 would contain $10^{-16}$ free-floating planets or brown dwarfs. The available photometry covers around 1000 times the duration of the eclipse. Thus, the chances of finding an eclipse in that observing span are in the range of $10^{-13}$ and thus negligible. We conclude that this explanation is not plausible.

\section{Summary and outlook}

We report the discovery of a deep, unique eclipse in the brown dwarf Roque 12, a bona fide member of the young cluster Pleiades with an age of 120\,Myr. The eclipse was observed in 2002 with two telescopes simultaneously, is 0.65\,mag deep and lasted 1.0-1.6 hours. The Kepler/K2 lightcurve from 2015 does not show any signs of further eclipses, ruling out that the event is periodic with periods shorter than 70\,d. There is tentative evidence for radial velocity variations in this target of $\sim 5$\,kms$^{-1}$. 

It is in principle conceivable that the eclipse was caused by a circum-sub-stellar dust cloud. Obscuration by objects along the line of sight, but unrelated to Roque 12 are considered highly unlikely. The most plausible explanation, however, is the presence of a sub-stellar companion on an eccentric orbit, with $e>0.5$. A low mass brown dwarf or a giant planet as companion fit the observational constraints, with a mass ratio $q<0.7$. The Roque 12 system could be one of very few known eclipsing binaries in the brown dwarf domain, the first with high eccentricity and long period. A system like that would be uniquely suited to test evolutionary models for sub-stellar objects. 

What are the prospects of confirming the presence of a companion? Apart from improved radial velocity constraints (see Sect. \ref{rv}), astrometry from Gaia may in the near future be able to provide more information. For the plausible system parameters (see Sect. \ref{companion}) the semi-minor axis times two corresponds to 0.5 to 4\,mas in the plane of the sky. For the semi-major axis times two, it would be up to 8\,mas. In Gaia DR2, the astrometric excess noise listed for R12 is nearly 4\,mas, and thus already excludes parts of the possible orbital parameter space.

Further photometric monitoring would allow us to either fix the quantity $n\times P$ if a period is detected, or at least rule out a range of periods. This does require deep continuous observations at high cadence and is therefore not an easy task. R12 is too far north for the baseline survey of the Vera Rubin Observatory, but since it is only 4\,deg away from the ecliptic, the object may still be covered regularly as part of surveys for solar system objects, albeit at high airmass. In a best case scenario with one eclipse every three months, the odds to catch the object in eclipse with a single visit are about 1/2000 or 0.05\%. We encourage the astronomical community to help hunting for the putative second eclipse of Roque 12.

\section{Acknowledgements}

We thank the anonymous referee for a speedy and constructive report with comments that helped to improve the paper.
Arno Riffeser and the staff at Calar Alto supported the observations in 2002. Discussions with Martin Kuerster, Veselin Kostov, and Dawn Peterson were helpful in preparing this paper. The UKIDSS project is defined in \citet{2007MNRAS.379.1599L}; the pipeline processing and science archive are described in \citet{2008MNRAS.384..637H}.  AS acknowledges support through
STFC grant ST/R000824/1. KM acknowledges funding by the Science and Technology Foundation of Portugal (FCT), grants No. IF/00194/2015, PTDC/FIS-AST/28731/2017 and UIDB/00099/2020. The Wetton 2018 workshop ‘Planning for Surprises - Data Driven Discovery in the era of Large Data’ in Oxford provided the incentive for publishing the paper.

\end{document}